\documentclass[]{pasj02} 
\usepackage[switch,mathlines]{lineno} 
\usepackage{url}
\usepackage{fix-cm}
\usepackage{xcolor}  

\jyear{2026}
\Received{}
\Accepted{}

\begin{document} 

\title{The Eclipsing $\gamma$ Doradus Star VZ Canum Venaticorum}

\author{Jae Woo \textsc{Lee}\altaffilmark{1}\altemailmark\orcid{0000-0002-5739-9804}} 
\email{jwlee@kasi.re.kr}

\altaffiltext{1}{Korea Astronomy and Space Science Institute, Daejeon 34055, Republic of Korea}


\KeyWords{asteroseismology --- binaries: eclipsing --- stars: fundamental parameters --- stars: individual (VZ Canum Venaticorum) --- stars: oscillations (including pulsations)}{}

\maketitle

\begin{abstract}
We present a detailed analysis of the eclipsing binary VZ CVn using TESS Sectors 50 and 77 photometry together with 
published spectroscopic results. The TESS light curves were modeled to refine the binary parameters, and 
the residual light curves were analyzed to investigate the pulsational properties of the system. We derive masses 
and radii of $M_{\rm A}=1.526\pm0.005$ $M_\odot$ and $R_{\rm A}=1.638\pm0.023$ $R_\odot$ for the primary and 
$M_{\rm B}=1.259\pm0.005$ $M_\odot$ and $R_{\rm B}=1.133\pm0.016$ $R_\odot$ for the secondary. Comparison with 
PARSEC evolutionary models suggests a near-solar metallicity of $Z=0.017$ and an age of 1.0 Gyr. 
Separate frequency analyses of the two TESS sectors reveal four signals with possible counterparts in both sectors. 
Two of these frequencies, $f_2=4.7486$ day$^{-1}$ and $f_6=7.1242$ day$^{-1}$, are consistent with the fourth and 
sixth orbital harmonics, respectively, suggesting an orbital origin rather than independent $\delta$ Sct pulsations. 
The dominant frequency $f_1=1.7577$ day$^{-1}$ has a pulsation constant of $Q_1=0.336$ day, consistent with 
a $\gamma$ Dor-type g-mode pulsation. The weaker signal $f_5=1.9754$ day$^{-1}$ also has a pulsation constant of 
$Q_5=0.299$ day, although its detection remains tentative because of its low S/N. Together with the physical and 
evolutionary properties of the primary, these results support its classification as a $\gamma$ Dor pulsator rather 
than the previously proposed $\gamma$ Dor--$\delta$ Sct hybrid pulsator. 
\end{abstract}


\section{Introduction} \label{sec_intro}

Pulsating stars in eclipsing binary (EB) systems allow their pulsational properties to be examined together with 
independently determined stellar parameters. In particular, accurate masses and radii derived from eclipsing and 
double-lined spectroscopic binaries provide strong constraints on the structure and evolutionary states of 
the pulsating components. These independently determined parameters are also useful for testing pulsation models and 
identifying the nature of the observed oscillations. Consequently, EBs containing pulsating components have received 
considerable attention in studies of stellar pulsation and evolution (see, e.g., \cite{Murphy2018,Southworth+2026}).

Despite the increasing number of pulsating stars identified in EBs, the pulsational nature of some systems remains uncertain 
or may require revision as improved observations become available. In particular, high-precision space photometry has made 
it possible to detect previously unresolved pulsation signals and to reassess earlier classifications. For example, GK Dra 
had long been known to exhibit $\delta$ Sct pulsations, but an analysis of TESS photometry revealed additional low-frequency 
signals attributable to $\gamma$ Dor pulsations, indicating the presence of both $\delta$ Sct and $\gamma$ Dor pulsations 
in the system \citep{Southworth2024}. Similarly, the primary of V421 Peg, previously proposed as a $\gamma$ Dor candidate, 
was recently confirmed as a $\gamma$ Dor pulsator using TESS photometry \citep{Lee2025}. These examples illustrate 
the importance of detailed frequency analyses for establishing the pulsational characteristics of EBs. 

VZ CVn (TIC 368129261, TYC 2003-780-1, HIP 66017, Gaia DR3 1449817995186764288) is a bright, double-lined EB with 
an orbital period of about 0.84 day. Its variability was discovered by \citet{Strohmeier+1960}, and the system was subsequently 
classified as a $\beta$ Lyr-type EB by \citet{Strohmeier+1962}. Early photometric observations and light-curve analyses were 
carried out by \citet{Harris1968}, \citet{Ibanoglu1974}, and \citet{Cester+1977}. The binary properties of VZ CVn became 
better constrained with the first spectroscopic orbit obtained by \citet{Popper1988}, who combined it with 
the previous photometric results to determine the physical properties of the components. \citet{Ibanoglu+2007} combined 
new two-color photometry with the radial-velocity (RV) results of \citet{Popper1988}, adopting the spectroscopic mass ratio 
of $q=0.777$, to derive revised absolute parameters for both components. More recently, \citet{Pribulla+2009} obtained 
a new double-lined spectroscopic orbit from DDO observations, yielding a more precise mass ratio of $q=0.8252\pm0.0019$. 
These spectroscopic results were used by \citet{Latkovic2012} in analyzing more than 14,000 WASP measurements, providing 
further refinements to the photometric and physical parameters of the system. 

In addition to its eclipsing variations, VZ CVn has long been known to exhibit intrinsic light variability. \citet{Ibanoglu1974} 
reported light variations with an amplitude of about 0.05 mag that could not be attributed to eclipses or other binary effects. 
The pulsational nature of this variability was later investigated by \citet{Ibanoglu+2007}, who extracted a dominant frequency 
of 0.93566 day$^{-1}$ (corresponding to a period of 1.06876 day) after removing the eclipsing and proximity effects from 
their light curves. Based on this period, the pulsation constant, and the position of the primary in the Hertzsprung--Russell 
(H-R) diagram, they identified the primary as a strong candidate for a $\gamma$ Dor pulsator. \citet{Latkovic2012} also 
searched the WASP residuals for pulsational variability and detected signals at 0.9659 and 1.7101 day$^{-1}$, 
although their significance was considered questionable because of strong daily aliasing. Most recently, \citet{Geldi+2025} 
analyzed ground-based and TESS photometry and reported four frequencies from their pulsation analysis. They interpreted two 
of these frequencies as $\gamma$ Dor pulsations and the other two as $\delta$ Sct pulsations, leading to the classification 
of the primary as a $\gamma$ Dor--$\delta$ Sct hybrid pulsator. This classification warrants further examination using 
the higher-time-resolution TESS Sector 77 observations now available for VZ CVn, which also allow a search for 
additional high-frequency pulsation signals. In this study, we analyze the TESS Sectors 50 and 77 photometry together with 
the spectroscopic results of \citet{Pribulla+2009} to refine the binary properties of VZ CVn and to re-examine the pulsational 
nature of its primary component.

\section{TESS Photometry and Eclipse Timings} \label{sec_photometry}

\begin{figure}
\begin{center}
\includegraphics[scale=0.9]{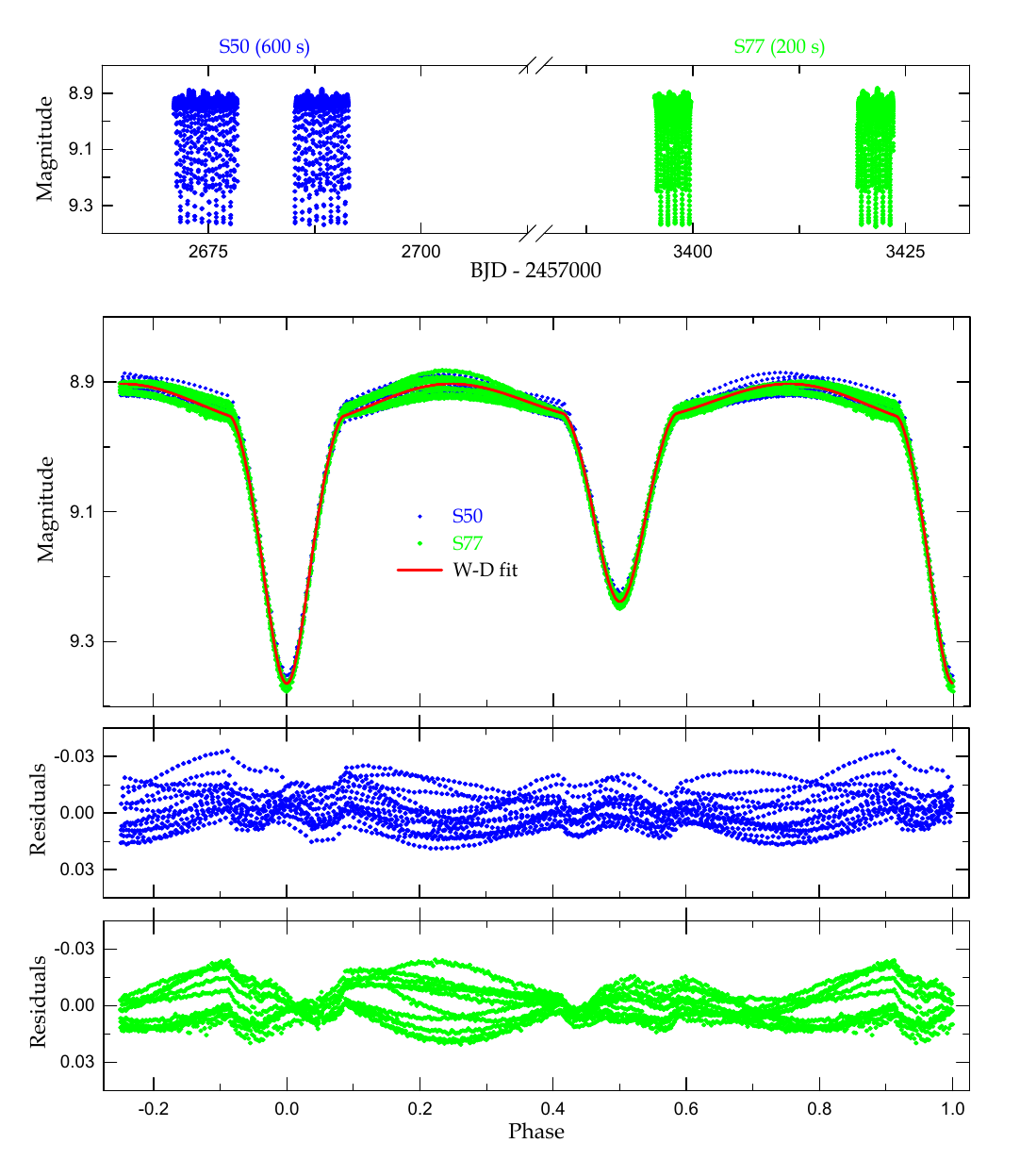}
\end{center}
\caption{TESS observations of VZ CVn distributed in BJD (top panel) and orbital phase (second panel). The blue and 
green circles are individual measurements for Sectors 50 and 77, respectively, and the red solid curve represents 
the synthetic model obtained through our W-D fit. The third and bottom panels show the corresponding residuals from  
the binary model.  \\
{Alt text: TESS light curves of VZ CVn from Sectors 50 and 77 plotted against BJD and orbital phase, together with 
the synthetic binary model. The residuals after subtracting the model are shown in the lower two panels.} } 
\label{Fig1}
\end{figure}

VZ CVn was observed by TESS \citep{Ricker+2015} in Sectors 23 (S23), 50 (S50), and 77 (S77), with cadences of 1800 s, 600 s, 
and 200 s, respectively. The higher-cadence S50 and S77 observations were used for the main binary and pulsation analyses, 
while the S23 data were also examined to independently check the detected frequencies. The TESS data were downloaded from 
the MAST Portal\footnote{\url{https://mast.stsci.edu/portal/Mashup/Clients/Mast/Portal.html}}. 
We used the SAP$\_$FLUX data and applied the detrending and normalization procedure described by \citet{Lee+2024}. 
Each sector was divided into two segments at the internal data gap, and a quadratic polynomial was fitted independently to 
the out-of-eclipse portions of each segment. The normalized fluxes were then converted to magnitudes by adopting 
the TESS magnitude of $T_{\rm p}=8.903$ from the TESS Input Catalog (TIC) v8.2 \citep{Paegert+2022} as the maximum-light level. 
The resulting S50 and S77 light curves are displayed in the first and second panels of Figure~\ref{Fig1} as a function of BJD 
and orbital phase, respectively. As shown in this figure, the S77 observations contain a substantial data gap caused by 
a missed data downlink and the subsequent Safe Hold of the spacecraft, as reported in the TESS S77 Data Release Notes 
(DRN 108)\footnote{\url{https://archive.stsci.edu/tess/tess_drn.html}}. 

Eclipse timings were measured from the processed S50 and S77 light curves using the method of \citet{Kwee+1956}. 
The measured timings and their uncertainties are listed in the first and second columns of Table~\ref{Tab1}, respectively, while 
the eclipse types are given in the fifth column. To obtain an ephemeris appropriate for phasing the TESS observations, 
a linear fit was applied to the timing measurements. The resulting ephemeris is 
\begin{equation}
C = \mbox{BJD}~2460396.20693(14) + 0.84246160(21)E, 
\label{eq:ephem}
\end{equation}
where $E$ denotes the cycle number. The epochs and $O-C$ residuals computed with this ephemeris are given in the third and 
fourth columns of Table~\ref{Tab1}, respectively.

\section{Binary Modeling and Absolute Parameters} \label{sec_binarity}

The TESS light curves of VZ CVn from S50 and S77 were analyzed using the Wilson--Devinney (W-D) code \citep{Wilson+1971,Kallrath2022}, 
in the same manner as in our previous studies of detached EBs (e.g., \cite{Lee2025}). The morphology of the TESS light curves 
and previous studies indicate that VZ CVn is a detached binary, and hence the analysis was performed in mode 2 of the W-D code. 
Although an earlier spectroscopic orbit was obtained by \citet{Popper1988}, we adopted the more recent RV solution of \citet{Pribulla+2009}, 
derived from DDO observations using the broadening-function technique. Their RV semi-amplitudes of $K_{\rm A}=141.51\pm0.26$ km s$^{-1}$ 
and $K_{\rm B}=171.48\pm0.25$ km s$^{-1}$ yield a spectroscopic mass ratio of $q=K_{\rm A}/K_{\rm B}=0.8252\pm0.0019$, which was 
used in our binary modeling. 
 
\citet{Pribulla+2009} classified VZ CVn as F0V based on their classification spectra. Although this spectral type refers to 
the combined spectrum of the binary system, the primary was found to contribute approximately 80 \% of the total $V$-band light 
in previous light-curve analyses \citep{Ibanoglu+2007,Geldi+2025}. Therefore, the F0V classification is expected to be 
representative of the primary component. According to the spectral type--effective temperature calibration of \citet{Pecaut+2013}, 
F0V corresponds to $T_{\rm eff}=7220$ K. Since \citet{Pribulla+2009} did not provide an uncertainty for the spectral classification, 
we assumed an uncertainty of one spectral subclass, corresponding to approximately 200 K around F0V in the same calibration. 
We thus adopted $T_{\rm eff,A}=7220\pm200$ K for the primary. This value is also consistent, at approximately the $1\sigma$ level, 
with $T_{\rm eff,A}=7000\pm24$ K independently estimated by \citet{Ibanoglu+2007} from the $B-V$ color index. 

The logarithmic limb-darkening law was adopted, with the coefficients interpolated from the \citet{van1993} tables available 
in the W-D code. The bolometric albedos and gravity-darkening exponents were fixed at the standard values of $A_{\rm A}=1.0$ 
and $g_{\rm A}=1.0$ for the primary, and $A_{\rm B}=0.5$ and $g_{\rm B}=0.32$ for the secondary. Synchronous rotation was 
assumed for both components by setting $F_{\rm A}=F_{\rm B}=1.0$. 

The adjustable parameters were the orbital ephemeris ($T_0$, $P_{\rm orb}$), inclination angle ($i$), secondary temperature 
($T_{\rm eff,B}$), dimensionless potentials ($\Omega_{\rm A}$, $\Omega_{\rm B}$), and primary luminosity ($L_{\rm A}$). 
These parameters were iteratively adjusted using the differential-corrections program of the W-D code. 
During the fitting process, we also tested solutions allowing for orbital eccentricity ($e$) and third light ($l_3$), but 
neither parameter yielded a statistically significant nonzero value. We therefore adopted a circular orbit ($e=0$) with 
no third-light contribution ($l_3=0$) in the final binary solutions. 
The S50 and S77 light curves were analyzed both separately and as a combined data set. Because the individual solutions 
were consistent within their uncertainties, the combined solution was adopted and is presented in Table~\ref{Tab2}. 
The uncertainties in both the mass ratio and the primary effective temperature were taken into account in estimating the errors 
of the fitted binary parameters. Following the approach of \citet{Southworth+2022}, we assessed the parameter 
uncertainties by exploring alternative W-D solutions under different modeling assumptions. 

The light-curve parameters derived from our binary modeling were combined with the RV semi-amplitudes $K_{\rm A}$ and 
$K_{\rm B}$ of \citet{Pribulla+2009} to determine the absolute dimensions of VZ CVn. These quantities are summarized in 
the lower part of Table~\ref{Tab2}. The bolometric corrections (BCs) for the two components were calculated from 
their effective temperatures using the calibration of \citet{Flower1996} and the corrections provided by \citet{Torres2010}. 
The distance to VZ CVn was determined from the distance modulus using the apparent magnitude $V=9.24\pm0.03$ and color excess 
$E(B-V)=0.014\pm0.003$, with the interstellar extinction estimated as $A_V=3.1E(B-V)$. The $V$ and $E(B-V)$ values were 
taken from TIC v8.2 \citep{Paegert+2022}. The EB-based distance of $229\pm13$ pc is in excellent agreement with 
the distance of $227.1\pm0.8$ pc inferred from the Gaia DR3 parallax ($\pi=4.404\pm0.016$ mas; \cite{Gaia2023}). 

\begin{figure}
\begin{center}
\includegraphics[]{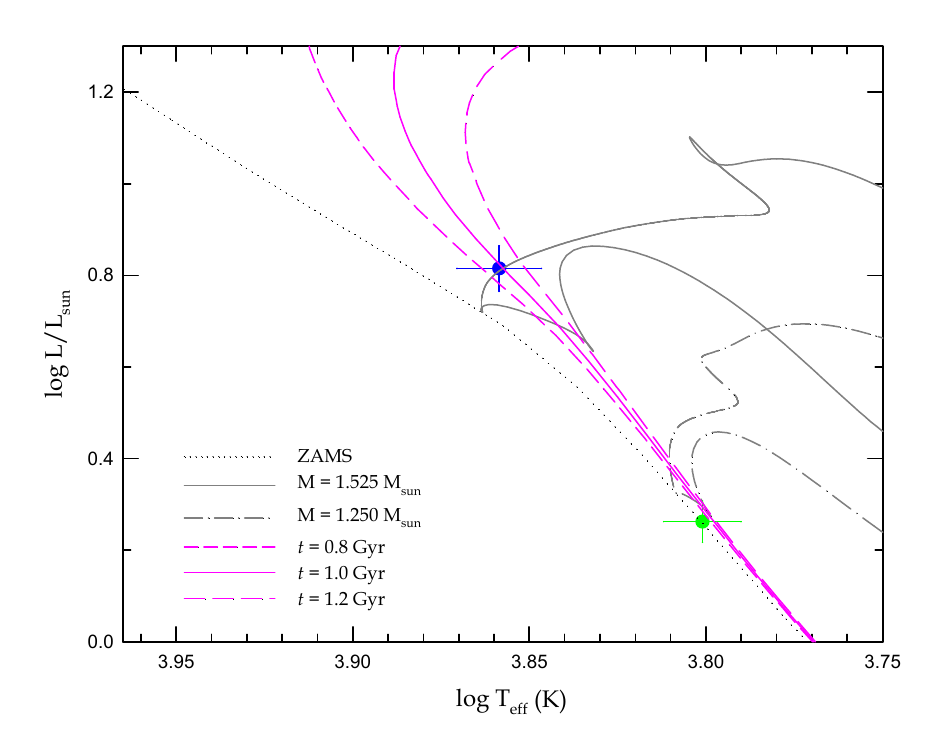}
\end{center}
\caption{H-R diagram of VZ CVn. The filled circles with error bars represent the primary and secondary components. 
The evolutionary tracks for 1.525 and 1.250 $M_\odot$, together with isochrones for ages of 0.8, 1.0, and 1.2 Gyr, 
are based on the PARSEC v1.2S models for $Z=0.017$ \citep{Bressan+2012}. The 1.525 $M_\odot$ track was obtained by 
interpolation between the 1.50 and 1.55 $M_\odot$ models. The ZAMS is shown by the dotted line. \\
{Alt text: H-R diagram showing the primary and secondary components of VZ CVn compared with PARSEC evolutionary tracks and 
isochrones. The primary appears to have evolved away from the ZAMS, while the secondary remains relatively close to it.} }
\label{Fig2}
\end{figure}

To investigate the evolutionary status of VZ CVn, we compared the observed properties of both components with the PARSEC v1.2S 
stellar evolutionary models \citep{Bressan+2012}. Figure~\ref{Fig2} shows the evolutionary tracks for masses close to those 
of the two components, together with isochrones for ages of 0.8, 1.0, and 1.2 Gyr. The observed positions of both components 
are broadly consistent with the models for a near-solar metallicity of $Z=0.017$ and an age of approximately 1.0 Gyr. 
The primary appears to have evolved away from the ZAMS, while the secondary remains relatively close to it.

\section{Pulsation Analysis} \label{sec_pulsation}

\begin{figure}
\begin{center}
\includegraphics[scale=0.9]{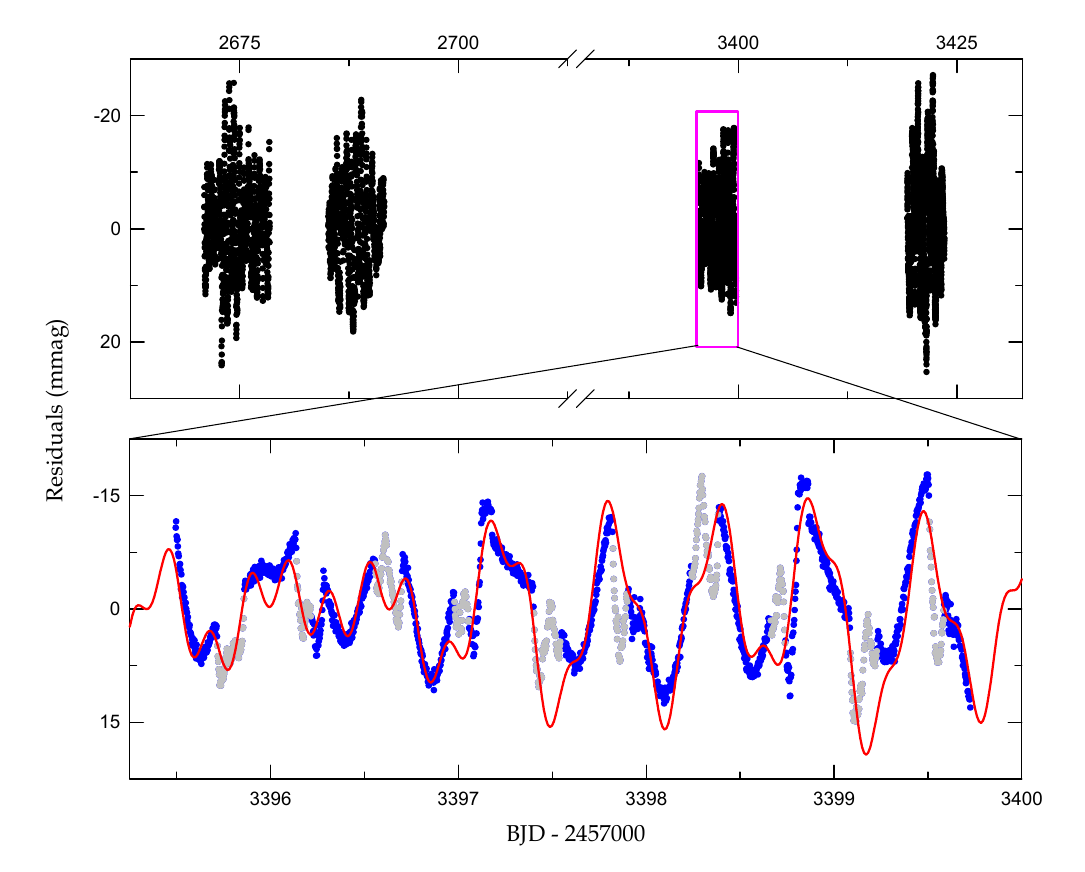}
\end{center}
\caption{Light curve residuals as a function of BJD. The lower panel shows an enlarged view of the boxed region 
in the upper panel. The blue and grey circles represent the out-of-eclipse and in-eclipse residuals, respectively, 
while the red solid line represents the multi-frequency fit to the out-of-eclipse data. \\
{Alt text: Light curve residuals of VZ CVn showing the out-of-eclipse and in-eclipse data. The enlarged lower panel shows 
the corresponding residuals together with the multi-frequency fit derived from the out-of-eclipse data.} }
\label{Fig3}
\end{figure}

The residual light curves displayed as a function of orbital phase in the third and fourth panels of Figure~\ref{Fig1}, as well 
as the time-series residuals shown in Figure~\ref{Fig3}, reveal clear pulsational variations with peak-to-peak amplitudes reaching 
approximately 40 mmag. To minimize the effects of eclipses on the pulsation signals, we used only the out-of-eclipse residuals 
at orbital phases 0.085--0.415 and 0.585--0.915 for the frequency analysis. The analysis was performed with the PERIOD04 program 
\citep{Lenz+2005} over the frequency range up to the Nyquist limit of each data set \citep{Lee+2014}.

The frequency analysis was carried out independently for the TESS S50 and S77 residual light curves because the two sectors 
are separated by approximately two years. The time spans of the residual data are 20.60 days for S50 and 28.06 days for S77, 
corresponding to formal Rayleigh frequency resolutions of $R_{\rm S50}=0.0485$ day$^{-1}$ and $R_{\rm S77}=0.0356$ day$^{-1}$, 
respectively. Following \citet{Loumos+1978}, we adopted $1.5R$ as a practical criterion for frequency resolution. 
Signals detected in both sectors were considered possible counterparts when their frequency differences were 
within this criterion. Given its longer time span and higher cadence, the S77 frequencies were adopted as representative values, 
while the amplitudes and phases were determined separately for the two sectors. 

\begin{figure}
\begin{center}
\includegraphics[]{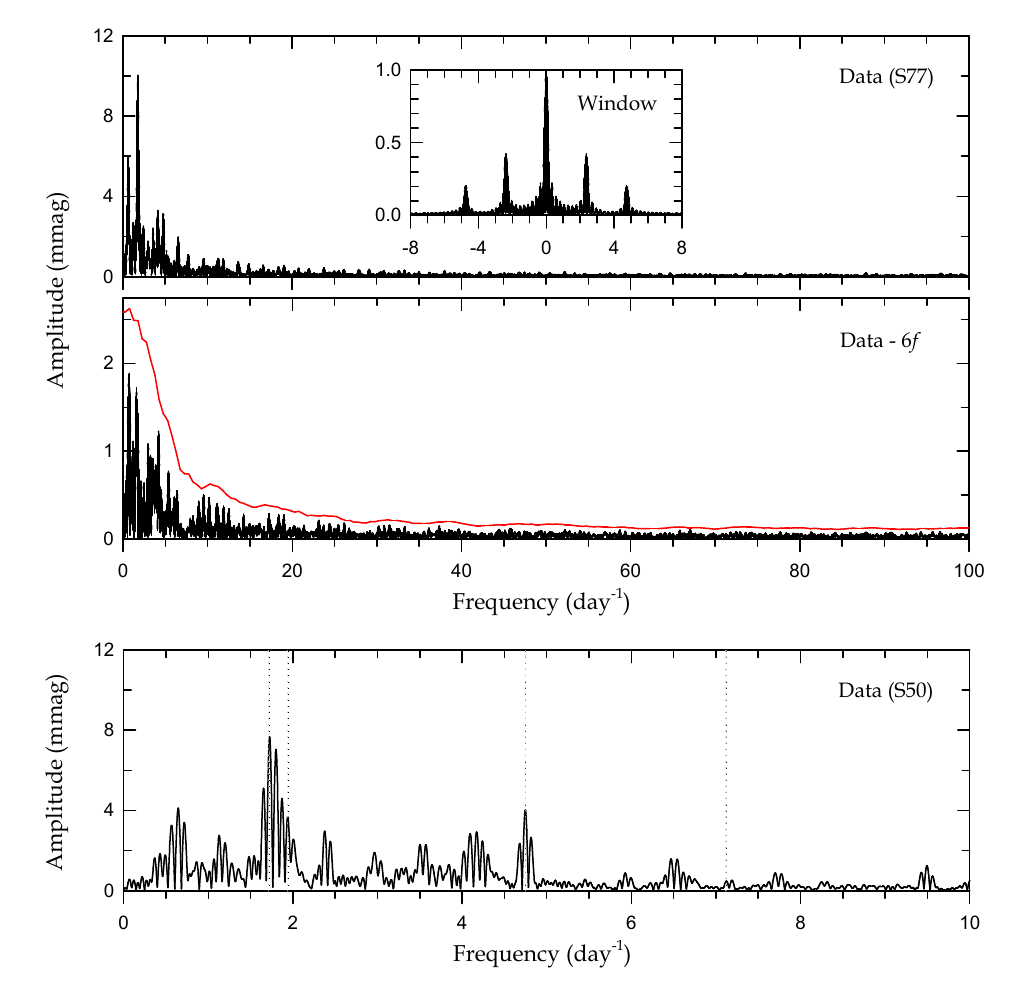}
\end{center}
\caption{PERIOD04 periodograms for the out-of-eclipse TESS residuals. The top and middle panels show the amplitude spectra 
of the 200-s S77 data before and after pre-whitening with the six detected frequencies, respectively. The inset in the top panel 
shows the spectral window function of the S77 data. The red solid line in the middle panel represents four times the noise level. 
The bottom panel shows the amplitude spectrum of the 600-s S50 data. The vertical dotted lines indicate the four frequencies 
reported by \citet{Geldi+2025}. \\{Alt text: PERIOD04 amplitude spectra of the TESS residuals of VZ CVn, showing the S77 data 
before and after pre-whitening with six detected frequencies, the corresponding spectral window function in the inset, and 
the S50 data for comparison. }} 
\label{Fig4}
\end{figure}

Using S/N $>4.0$ as the threshold for extracting candidate frequencies to facilitate comparison with the results of 
\citet{Geldi+2025}, we identified six frequencies in each sector. However, \citet{Baran+2021} showed that higher detection 
thresholds of S/N $\sim5$ may be appropriate for TESS photometry, depending on the observing cadence and data coverage. 
We therefore regard candidate frequencies with S/N below approximately 5 as tentative detections. Figure~\ref{Fig4} presents 
the amplitude spectra of the S77 and S50 data, together with the residual spectrum after pre-whitening the six S77 frequencies. 
The frequencies detected in S77 are listed in Table~\ref{Tab3} in the order of their detection and are designated $f_1$ through 
$f_6$. No additional significant frequencies were detected beyond the frequency ranges displayed in Figure~\ref{Fig4}.  

An additional frequency detected only in S50 is designated as $f_7$. Four frequencies, $f_1$, $f_2$, $f_5$, and 
$f_6$, have possible counterparts in both sectors. The frequencies of $f_1$ and $f_5$ differ between S50 and S77 by more than 
their formal uncertainties, and the amplitude of $f_1$ also varies between the two sectors. The origin of these differences remains 
uncertain, and the correspondence of the signals does not necessarily imply identical pulsation modes. However, the dominant frequency 
$f_1$ lies in the $\gamma$ Dor frequency domain in both sectors. 

The S77 frequency $f_3$, which was not detected in S50, lies close to the combination frequency $f_1-f_{\rm orb}$, with 
a difference of 0.0585 day$^{-1}$, suggesting a possible association. The signals designated as $f_4$ were detected in both sectors, 
although their frequencies, $f_{\rm S77}=1.2741$ day$^{-1}$ and $f_{\rm S50}=1.2008$ day$^{-1}$, differ by 0.0733 day$^{-1}$. 
This difference slightly exceeds $1.5R_{\rm S50}$, and we therefore regard the association of the two $f_4$ signals 
as tentative. For S77, the frequency difference between $f_4$ and $f_1-f_3$ is 0.0286 day$^{-1}$, which is smaller than 
the adopted resolution criterion. In contrast, the S50 frequency of $f_4$ is close to the orbital frequency, with a difference 
of only 0.0138 day$^{-1}$, suggesting that the two $f_4$ signals may have different origins. \citet{Geldi+2025} reported 
four frequencies at 1.71273 day$^{-1}$, 4.74934 day$^{-1}$, 1.937549 day$^{-1}$, and 7.120613 day$^{-1}$ from their analysis 
of the TESS data. These frequencies can be associated with our $f_1$, $f_2$, $f_5$, and $f_6$, respectively. 
In particular, the frequencies of $f_2$ and $f_6$, which are close to the fourth and sixth orbital harmonics, agree well with 
their results. In our analysis, only the out-of-eclipse residuals were used, whereas the treatment of eclipse phases in 
the previous study was not explicitly specified. Differences in data selection or frequency-analysis procedures may partly 
explain the discrepancies between the two studies. 

Although frequencies coinciding with orbital harmonics may also arise from tidally excited or perturbed pulsations, such oscillations 
are generally expected to be less prominent in circular binaries than in eccentric systems. The spectral window function shown in 
Figure~\ref{Fig4} exhibits structures at even multiples of the orbital frequency, reflecting the periodic masking of both eclipses. 
These findings support an association of $f_2$ and $f_6$, corresponding to $4f_{\rm orb}$ and $6f_{\rm orb}$, respectively, with 
orbital variability rather than independent $\delta$ Sct pulsations. 

For an independent comparison, we also analyzed the TESS S23 observations using the binary parameters derived 
in this study. The frequency analysis of the out-of-eclipse residuals yielded three signals with S/N $>5$ at 1.9380, 1.7090, and 
1.8051 day$^{-1}$. The first two frequencies are close to those identified as $f_5$ and $f_1$, respectively, in S50 and S77. 
In contrast, no significant signals corresponding to $f_2$ and $f_6$, which lie close to $4f_{\rm orb}$ and $6f_{\rm orb}$, 
respectively, were detected in S23. The S23 residual data span 24.60 days, corresponding to a formal Rayleigh frequency resolution 
of $R_{\rm S23}=0.0407$ day$^{-1}$.

\section{Summary and Discussion} \label{sec_summary}

We have investigated VZ CVn using the TESS S50 and S77 photometry together with the spectroscopic results of \citet{Pribulla+2009}. 
The TESS light curves were analyzed to refine the binary parameters and to investigate the pulsational characteristics of 
its primary component. The main results of this study are summarized as follows. 

\begin{enumerate}
\item The TESS photometric data were modeled with the W-D code by adopting the spectroscopic mass ratio of $q=0.8252\pm0.0019$ 
and an effective temperature of $T_{\rm eff,A}=7220\pm200$ K for the primary, estimated from its F0V spectral type. The solutions 
obtained independently from S50 and S77 were consistent within their uncertainties, and the combined solution was adopted 
to determine the absolute dimensions of the system. The resulting masses and radii are $M_{\rm A}=1.526\pm0.005$ $M_\odot$, 
$M_{\rm B}=1.259\pm0.005$ $M_\odot$, $R_{\rm A}=1.638\pm0.023$ $R_\odot$, and $R_{\rm B}=1.133\pm0.016$ $R_\odot$. 
The EB-based distance of $229\pm13$ pc agrees excellently with the Gaia DR3 distance of $227.1\pm0.8$ pc. Comparison with 
the PARSEC v1.2S models suggests a near-solar metallicity of $Z=0.017$ and an age of approximately 1.0 Gyr. 

\item Independent frequency analyses of the S50 and S77 out-of-eclipse residuals revealed six candidate frequencies with 
S/N $>4.0$ in each sector. Among them, $f_1$, $f_2$, $f_5$, and $f_6$ have possible counterparts in both sectors.  
For these signals, the S77 frequencies were adopted as the representative values. The remaining signals include $f_3$, 
possibly associated with $f_1-f_{\rm orb}$, and the S77 signal $f_4$, possibly associated with $f_1-f_3$. The S50 signal 
$f_4$ lies close to the orbital frequency, while $f_7$ was detected only in S50.  

\item \citet{Geldi+2025} interpreted their first (1.713 day$^{-1}$) and third (1.938 day$^{-1}$) frequencies as 
$\gamma$ Dor pulsations and their second (4.749 day$^{-1}$) and fourth (7.121 day$^{-1}$) frequencies as $\delta$ Sct 
pulsations, and consequently classified the VZ CVn primary as a $\gamma$ Dor--$\delta$ Sct hybrid pulsator. Our analysis, 
however, shows that $f_2$ and $f_6$ are consistent with $4f_{\rm orb}$ and $6f_{\rm orb}$, respectively. 
These signals are therefore more likely associated with orbital harmonics than with independent $\delta$ Sct  pulsations.
The dominant frequency $f_1$ lies in the $\gamma$ Dor frequency domain, with a pulsation constant of $Q_1=0.336$ day, consistent 
with a $\gamma$ Dor-type g-mode pulsation ($f\lesssim4$ day$^{-1}$ and $Q>0.23$ day; \cite{Handler+Shobbrook2002,Uytterhoeven+2011}). 
The weaker signal $f_5$, detected in both sectors, also lies in this frequency domain and has a pulsation constant of $Q_5=0.299$ day, 
although its detection remains tentative because of its low S/N. We therefore conclude that the primary component of VZ CVn 
is best classified as a $\gamma$ Dor pulsator, with no convincing evidence for independent $\delta$ Sct pulsations. 
\end{enumerate}

As an eclipsing binary containing a $\gamma$ Dor pulsator, VZ CVn provides an opportunity to study stellar pulsation with 
independently constrained fundamental parameters. Our results also highlight the importance of carefully interpreting signals 
at orbital harmonics when investigating stellar pulsations in close binary systems.  

\begin{ack} 
We thank the anonymous referee for the careful review and constructive comments that helped improve the manuscript.  
This paper has made use of the public TESS photometry and the SIMBAD database, operated at CDS, Strasbourg, France. 
We gratefully acknowledge the support from the KASI grant 2026-1-904-01. 
\end{ack}

\onecolumn{}
\clearpage

\clearpage 
\begin{table}
\tbl{TESS Eclipse Timings of VZ CVn. }{%
\begin{tabular}{lcccc}
\hline
BJD               & Error           & $E$           & $O$--$C$           & Min         \\ 
\hline
2,459,671.26832  & $\pm$0.00006	   & $-860.5$        & $-0.00040$      & II            \\
2,459,671.68977  & $\pm$0.00001	   & $-860.0$        & $-0.00018$      & I             \\
2,459,672.11089  & $\pm$0.00007	   & $-859.5$        & $-0.00030$      & II            \\
2,459,672.53163  & $\pm$0.00008	   & $-859.0$        & $-0.00079$      & I             \\
2,459,672.95420  & $\pm$0.00011	   & $-858.5$        & $+0.00055$      & II            \\
2,459,673.37662  & $\pm$0.00021	   & $-858.0$        & $+0.00174$      & I             \\
2,459,673.79552  & $\pm$0.00015	   & $-857.5$        & $-0.00059$      & II            \\
2,459,674.21620  & $\pm$0.00017	   & $-857.0$        & $-0.00114$      & I             \\
2,459,674.63871  & $\pm$0.00005	   & $-856.5$        & $+0.00014$      & II            \\
2,459,675.06071  & $\pm$0.00014	   & $-856.0$        & $+0.00091$      & I             \\
\hline
\end{tabular}}\label{Tab1}
\begin{tabnote}
Note: This table is available in its entirety in machine-readable form. A portion is shown here for guidance regarding its form and content.
\end{tabnote}
\end{table}

\begin{table}
\tbl{Binary Parameters of VZ CVn. }{%
\begin{tabular}{lcc}
\hline
Parameter                         & Primary (A)           & Secondary (B)               \\ 
\hline
$T_0$ (BJD)                       & \multicolumn{2}{c}{2,460,396.20691$\pm$0.00010}     \\   
$P_{\rm orb}$ (day)               & \multicolumn{2}{c}{0.84246161$\pm$0.00000014}       \\   
$q$                               & \multicolumn{2}{c}{0.8252$\pm$0.0019}               \\   
$i$ (deg)                         & \multicolumn{2}{c}{80.683$\pm$0.035}                \\   
$T_{\rm eff}$ (K)                 & 7220$\pm$200          & 6318$\pm$160                \\   
$\Omega$                          & 4.108$\pm$0.040       & 4.958$\pm$0.050             \\   
$\Omega_{\rm in}$$\dagger$        & \multicolumn{2}{c}{3.460}                           \\   
$x$, $y$                          & 0.518, 0.277          & 0.567, 0.270                \\   
$l/(l_{\rm A}+l_{\rm B})$         & 0.7538$\pm$0.0018     & 0.2462                      \\   
$r$ (pole)                        & 0.3014$\pm$0.0038     & 0.2120$\pm$0.0029           \\   
$r$ (point)                       & 0.3275$\pm$0.0055     & 0.2192$\pm$0.0034           \\   
$r$ (side)                        & 0.3093$\pm$0.0042     & 0.2143$\pm$0.0030           \\   
$r$ (back)                        & 0.3199$\pm$0.0049     & 0.2178$\pm$0.0033           \\   
$r$ (volume)$\ddagger$            & 0.3105$\pm$0.0043     & 0.2148$\pm$0.0031           \\ [1.0mm]
\multicolumn{3}{l}{Absolute dimensions:}                                                \\   
$M$ ($M_\odot$)                   & 1.526$\pm$0.005       & 1.259$\pm$0.005             \\   
$R$ ($R_\odot$)                   & 1.638$\pm$0.023       & 1.133$\pm$0.016             \\   
$\log$ $g$ (cgs)                  & 4.193$\pm$0.012       & 4.429$\pm$0.013             \\   
$\rho$ ($\rho_\odot$)             & 0.348$\pm$0.015       & 0.867$\pm$0.038             \\   
$L$ ($L_\odot$)                   & 6.53$\pm$0.75         & 1.83$\pm$0.19               \\   
$M_{\rm bol}$ (mag)               & 2.70$\pm$0.12         & 4.08$\pm$0.11               \\   
BC (mag)                          & 0.03$\pm$0.01         & $-0.01\pm0.02$              \\   
$M_{\rm V}$ (mag)                 & 2.67$\pm$0.12         & 4.09$\pm$0.12               \\   
Distance (pc)                     & \multicolumn{2}{c}{229$\pm$13}                      \\   
\hline
\end{tabular}}\label{Tab2}
\begin{tabnote}
\footnotemark[$\dagger$]Potential for the inner critical Roche surface. \\
\footnotemark[$\ddagger$]Mean volume radius. 
\end{tabnote}
\end{table}

\begin{table}
\tbl{Frequency solutions for VZ CVn from TESS Sectors 50 and 77. }{%
\begin{tabular}{lcccccccc}
\hline
         & $f_{\rm S77}$       & $A_{\rm S77}$       & S/N$_{\rm S77}$       & $f_{\rm S50}$       & $A_{\rm S50}$       & S/N$_{\rm S50}$       & $f_{\rm S77}-f_{\rm S50}$  & Identification                  \\
\hline                                                                                         
$f_{1}$  & 1.7577(07)          & 10.4(1.1)           & 16.73                 & 1.7231(12)          & 7.9(1.0)            & 13.90                 & $+0.0346$                  &                                 \\
$f_{2}$  & 4.7486(10)          &  4.2(0.6)           & 11.81                 & 4.7496(11)          & 4.7(0.6)            & 14.47                 & $-0.0010$                  & $4f_{\rm orb}$                  \\
$f_{3}$$\dagger$ & 0.5122(28)  &  2.8(1.1)           &  4.33                 & \dots               & \dots               & \dots                 & \dots                      & $f_{1}-f_{\rm orb}$, S77 only   \\
$f_{4}$$\ddagger$ & 1.2741(28) &  2.7(1.1)           &  4.37                 & 1.2008(30)          & 3.2(1.0)            &  5.48                 & $+0.0733$                  & $f_{1}-f_{3}$                   \\
$f_{5}$  & 1.9754(29)          &  2.4(1.0)           &  4.16                 & 1.9989(38)          & 2.4(1.0)            &  4.34                 & $-0.0235$                  &                                 \\
$f_{6}$  & 7.1242(12)          &  1.9(0.3)           & 10.26                 & 7.1245(12)          & 2.1(0.3)            & 13.92                 & $-0.0003$                  & $6f_{\rm orb}$                  \\
$f_{7}$  & \dots               & \dots               & \dots                 & 1.8745(22)          & 4.3(1.0)            &  7.52                 & \dots                      & S50 only                        \\
\hline
\end{tabular}}\label{Tab3}
\begin{tabnote}
Note: Frequencies ($f_{\rm S77}$, $f_{\rm S50}$) and amplitudes ($A_{\rm S77}$, $A_{\rm S50}$) are given in units of day$^{-1}$ and mmag, respectively. The numbers in parentheses are the uncertainties in the last significant digits, estimated following \citet{Kallinger+2008}. Frequencies are listed in the order of their detection in S77; $f_{7}$, detected only in S50, is listed last.  \\
\footnotemark[$\dagger$]The identification of $f_3$ as $f_{1}-f_{\rm orb}$ is tentative because the frequency difference of 0.0585 day$^{-1}$ slightly exceeds the adopted resolution criterion of 1.5$R_{\rm S77}$=0.0534 day$^{-1}$. \\
\footnotemark[$\ddagger$]The association of the S50 and S77 signals for $f_{4}$ is tentative because their frequency difference is approximately 1.5$R_{\rm S50}$. The S77 frequency of $f_4$ is consistent with $f_1-f_3$ within the adopted resolution criterion, whereas the S50 frequency is 
close to $f_{\rm orb}$. 
\end{tabnote}
\end{table}

\end{document}